\documentstyle[prd,aps]{revtex} 
\tightenlines
\global\arraycolsep=2pt 
\begin{document}
\title{The exact solution for the fluorescence of low density 
Frenkel excitons in double and triple lattice-layers}
\author{Chang-qi Cao$^{a,b}$\footnote{email:cqcao@imech.ac.cn },  
Liu Yu-Xi$^{b,c}$\footnote{email:liuyx@itp.ac.cn} and 
Hui Cao$^{d}$\footnote{email:h-cao@nwu.edu}}
\address{a. CCAST(World Lab), P. O. Box 8730, Beijing 10080, China}
\address{b. Department of Physics,  Peking University,  Beijing 100871,  China}
\address{c. Institute of Theoretical Physics, Academia Sinica, 
P.O.Box 2735, Beijing 100080, China}
\address{d. Department of Physics and Astronomy,  Northwestern University,Evanston,  
IL 60208--3112 USA}
\maketitle

\begin{abstract}
 In low density regime , the fluorescence of Frenkel exitons in crystal slab
 can be studied without the aid of rotating wave and Marckoffian approximation. 
 The equations for the case of double and triple lattice-layers are now solved 
 exactly to give the eigen decay rates, frequency shifts and the statistical
  properties of the fields.  
 \end{abstract}
\vspace{1cm}
\hspace{1.2cm}
{PACS number(s): 42.50 Fx,  71.35-y}

\section {Introduction }
The fluorescence of excitons in a quantum well or 
crystal slab is of 
charater of collective radiation, since exciton, as exited state 
of the whole quantum
well (crystal slab) has a collective transition dipole moment.  
However not all these collective radiation are superradiance. 
Actually, the exciton emission has many eigen modes, some of them are 
superradiant modes and some of them are subradiant modes.

Exciton have important application in photonic devices because excitonc
device may have small size, low power dissipation, high speed and high 
efficiency.  All of these are needed by integrated photo-electric circuits.  

It is well known that the exciton in bulk crystal does not radiate, 
but forms polariton
instead$^{[1,2]}$.  This shows that a general treatment of exciton 
radiation should take the reabsorption effect into account.  

There has been quite a lot of theoretical studies on fluorescence  of
 excitons$^{[3-10]}$.  In the case of Frenkel 
excitons of low density, Knoester studied$^{[5]}$ the crossover from 
superradiant excitons to bulk polaritons when the number of the lattice-layers
in the crystal increases without bound.   He gives the correct form of the
eigen equation for the frequency shift and decay rate in the single 
lattice layer case. We have pointed out$^{[11]}$ that it should count 
in the two-photon coupling term properly in order to get this correct form of 
eigen equation.  Knoester proposed$^{[5]}$ that $F_{kk'}$ which describe the 
coupling between the excitons of wave vector $k$ and  $k'$ by exchange 
of photons is strongly peaked around $k=k'$ and one may keep only the diagonal 
elements to a good approximation.  There are some ambiguities in this 
proposition, since the set of  the values  for $k$ exists different selections.
Besides,  when one derives the whole set of eigen decay rates, some of them are 
small (subradiant modes), the contribution from off diagonal elements could be important.  
 
We have studied the single lattice layer case in some detail$^{11}$. However 
in the case of Frenkel exciton fluorescence, neither single layer case nor the
very thick case is important in practice. In this paper we will study the 
fluorescence of Frenkel exciton in 
thin crystal film of double and triple lattice-layers.  The  Heisenberg 
equations without rotating wave approximation are solved in the low density 
 regime without Mackoffian approximation.  The two-photon coupling term in the
 interaction Hamiltonian is included properly.  All eigen decay rates,
frequency shifts as  well as the evolution of fields in terms of their initial
values are obtained consequently.  We note the non-diagonal elements of 
$F_{kk'}$ is essential to derivation of these results. Our approach
can be readily generalized to the case of more lattice layers.
 
 The exciton-phonon interaction is not taken into account in this 
 investigation.  We shall give a brief review of general formulation in 
 section 2. Section 3 and 4 are devoted to the cases of double and triple 
 lattice-layers respectively. A brief conclusion is given in section 5.
 
 \section {Brief review of general formulation}
 
 We first write down the general formulation for the crystal slab of $N$
  lattice-layers.   The crystal is assumed to have simple cubic structure.  
  When the two-photon coupling  term $\frac{e^{2}}{2mc}A^{2}$ is taken into 
  account, the interaction Hamiltonian between exciton and photon for the low 
  density excitons are described as$^{[11]}$
\begin{eqnarray}
  \hat{H}_{int} & = & \hbar \sum_{q, k}G(q){\Large O} (k+q)
  [\hat{B}_{k}(t)+\hat{B}_{-k}^{+}(t)]
  [\hat{a}_{q}(t)+\hat{a}^{+}_{-q}(t)]+  \nonumber \\
                & & \hbar \sum_{q,q',k}\frac{1}{\Omega}G(q)G(q'){\Large O}
               (q'-k){\Large O}(k+q)[\hat{a}_{q}(t)+\hat{a}_{-q}^{+}(t)]
               [\hat{a}_{q'}(t)+\hat{a}_{-q'}^{+}(t)]
\end{eqnarray}
  where $q$ and $k$ are wave vectors for photon and exciton respectively, 
  they are in the $z$-direction, perpendicular to the crystal slab, 
  $\hat{B}_{k}(t)$ and 
  $\hat{a}_{q}(t)$ are exciton and photon annihilation operators respectively, 
  $\hat{B}_{k}(t)$ and $\hat{B}_{k}^{+}(t)$ are assumed to satisfy the boson 
  commutation relations 
\begin{equation}
  [\hat{B}_{k}(t) , \hspace{0.1cm}\hspace{0.3cm}\hat{B}_{k'}^{+}(t)]=
  \delta_{kk'}, 
\end{equation}
  and $\Omega$ is the frequency of the isolated lattice atom, $G(q)$ is the 
  coupling constant, 
\begin{equation}
  G^{2}(q)=\frac{2\pi N_{T}\Omega^{2}}{V\hbar |q|c} d^{2}
\end{equation}
 where $V=AL$ is the normalization volume for the photon, $N_{T}$ is the 
  total number of lattice sites in the crystal slab($N_{T}=N N_{L}$, $N_{L}$
 is the number of lattice sites in each layer). $d$ denotes the transition 
 dipole moments of a  single lattice atom supposed for simplicity perpendicular
  to the $z$-axis.

  In Ref.[5] it is said that $k$ takes the discrete value
  \begin{equation}
  k=\frac{2\pi m}{Na}
  \end{equation}
with $m=0, 1, \cdots N-1$.  But in the eq(1) of Ref.[5] as well as in our 
$\hat{H}_{int} $, $k$'s are assumed symmetric with respect to the $zero$, 
hence we should take $m=-\frac{1}{2}(N-1),
 \cdots, \frac{1}{2}(N-1)$ instead$^{[11]}$.  The values of $q$ is as usual:

$$q=\frac{2\pi j}{L}, \hspace{1cm}j=0, \pm 1, \pm 2, \cdots, \pm \infty.$$
${\Large O} (k-q)$ is the wave-vector matching factor, now takes the 
form$^{[11]}$

\begin{equation}
{\Large O} (k-q)=\frac{1}{N}\sum _{l=-\frac{1}{2}(N-1)}^{\frac{1}{2}(N-1)}
e^{i(k-q)la}=\frac{1}{N} \frac{\sin \frac{1}{2}(k-q)Na}{\sin\frac{1}{2}(k-q)a},
\end{equation}
$l$ is the index for the layers.  In case $k=q$, ${\Large O} (k-q)$ will 
equal to one, when $k\ne q$ and $N$ is sufficient large, ${\Large O}(k-q)$
 will be small.  

We notice that no rotating wave approximation is made in eq.(1), and the 
two terms on the right-hand side correspond to single-photon coupling and 
two-photon coupling respectively.  

From $\hat{H}_{int}$ and the commutation relation, one may immediately write 
down the Heisenberg equations for $\hat{B}_{k}(t)$, $\hat{a}_{q}(t)$ and 
their hermitian conjugates.  These equations are linear equations so that 
they can be solved exactly.  Carrying out the half-side Fourier transformation
\begin{equation}  
\hat{B}_{k}(\omega)=\int_{0}^{\infty}\hat{B}_{k}(t)e^{i\omega t}{\rm d}t,
\hspace{1cm} 
\hat{B}_{-k}^{+}(\omega)=\int_{0}^{\infty}\hat{B}_{-k}(t)e^{i\omega t}
{\rm d}t, \hspace{0.5cm}etc.  
\end{equation}
as did in Ref.[5] and then eliminating the photon operators, we get  
\begin{eqnarray}  
  & & \sum_{k'}[(\omega ^{2}-\Omega ^{2})\delta_{kk'}-
  \frac{2\omega^{2}}{\Omega} F_{kk'}(\omega)]
 [\hat{B}_{k}(\omega)-\hat{B}_{-k}^{+}(\omega)] \nonumber\\
 & = & i[(\omega +\Omega)\hat{B}_{k}(0)-(\omega -\Omega)\hat{B}_{-k}^{+}(0)]
 -2i\frac{\omega}{\Omega}\sum_{k'}F_{kk'}(\omega)[\hat{B}_{k'}(0)-\hat{B}_{-k'}^{+}(0)] \nonumber\\
 & & +2i\sum_{q}\omega G(q){\Large O} (k-q)[\frac{\hat{a}_{q}(0)}{\omega
 -|q|c}+\frac{\hat{a}_{-q}^{+}(0)}{\omega +|q|c}],
\end{eqnarray}

where $\hat{B}_{k'}(0)$ means $\hat{B}_{k'}(t=0)$, etc,  and 
\begin{eqnarray}  
F_{kk'}(\omega) & = & \sum_{q} \frac{2|q| c}{\omega^{2}-q^{2}c^{2}}G^{2}(q)
{\Large O} (k-q) {\Large O} (k'-q)\nonumber\\
 & = & -\frac{Na\Omega f^{2}}{4\pi c^{2}}\int {\rm d}q 
\frac{{\Large O} (k-q) {\Large O} (q-k')}{q^{2}-\omega^{2}/c^{2}}, 
\end{eqnarray}
in which
\begin{equation} 
f^{2}=\frac{8\pi\Omega d^{2}}{\hbar a^{3}}.
\end{equation}
Note that our $F_{kk'}(\omega)$ and ${\Large O}(k-q)$ are somewhat different 
from  those defined in Re[5].  For simplicity we have negelected the term 
of static dipole-dipole interactions 

We should mention that even disregarding the terms proportional to $\hat{a}
_{q}(0)$ and $\hat{a}_{-q}(0)$, eq.(7) is still different  from the 
result of Ref.[5] as mentioned in Ref.[11].

\section{The case of double lattice-layers}
Now we consider the special case $N=2$. The wave-vector matching factor for 
$N=2$ becomes
\begin{equation}
{\Large O} (k-q)=\cos \frac{1}{2} (k-q)a
\end{equation}
where $k$ takes the values $\pm \frac{\pi}{2a}$. We shall use $F_{++}(\omega)$,
$F_{--}(\omega)$ to represent $F_{+\frac{\pi}{2a}, +\frac{\pi}{2a}}(\omega)$,
 $F_{-\frac{\pi}{2a},-\frac{\pi}{2a}}(\omega)$ etc. and evaluate the 
 integrals in eq.(8) by contour integration. The results are
\begin{equation}  
F_{++}(\omega)=F_{--}(\omega)=-i\frac{\eta \Omega}{\omega},\hspace{1cm}
F_{+-}(\omega)=F_{-+}(\omega)
=-i\frac{\eta\Omega}{\omega} e^{\frac{i\omega a}{c}}
\end{equation}
where 
\begin{equation}  
\eta =\frac{af^{2}}{2c}.
\end{equation} 
We note that the nondiagonal elements  ($F_{+-}$ and $F_{-+}$) is of the 
same orderly as the diagonal elements ($F_{++}$ and $F_{--}$).

The coupled equation(7) now becomes 
\setcounter{equation}{0}
\def\theequation{13\alph{equation}}
\begin{eqnarray}  
(\omega^{2} +i\omega \eta -\Omega^{2})[\hat{B}_{+}(\omega)-\hat{B}_{-}^{+}
(\omega)]
+i\omega \eta e^{\frac{i\omega a}{c}}[\hat{B}_{-}(\omega)-\hat{B}_{+}^{+}
(\omega)] & = & \hat{A}_{0}(w),  \\
i\omega \eta e^{\frac{i\omega a}{c}}[\hat{B}_{+}(\omega)-\hat{B}_{-}^{+}
(\omega)] +(\omega^{2} +i\omega\eta -\Omega^{2})[\hat{B}_{-}(\omega)
-\hat{B}_{+}^{+}(\omega)]
 & = & -\hat{A}_{0}^{+}(-\omega)
\end{eqnarray}
where 
\begin{eqnarray}
\hat{A}_{0}(\omega) & = & i[(\omega +\Omega +i\eta)\hat{B}_{+}(0)-(\omega -
\Omega +i\eta)\hat{B}_{-}^{+}(0)]-\eta e^{\frac{i\omega a}{c}}[\hat{B}_{-}(0)-
\hat{B}_{+}^{+}(0)]\nonumber\\
 & & +2\sqrt{2}\omega i\sum_{q}G(q)\cos (\frac{\pi}{4}-\frac{qa}{2})
 [\frac{\hat{a}_{q}(0)}{\omega -|q|c}+\frac{\hat{a}_{-q}^{+}(0)}
 {\omega +|q|c}],\\
 \hat{A}_{0}^{+}(-\omega) & = & i[(\omega -\Omega +i\eta)\hat{B}_{+}^{+}(0)-
(\omega +\Omega +i\eta)\hat{B}_{-}(0)]-\eta e^{\frac{i\omega a}{c}}
[\hat{B}_{-}^{+}(0)-\hat{B}_{+}(0)]\nonumber\\
 & & -2\sqrt{2}\omega i\sum_{q}G(q)\cos (\frac{\pi}{4}+\frac{qa}{2})
 [\frac{\hat{a}_{q}(0)}{\omega -|q|c}+\frac{\hat{a}_{-q}^{+}(0)}{
 \omega +|q|c}]. 
\end{eqnarray}
\def\theequation{\arabic{equation}}
\setcounter{equation}{0}
\def\theequation{14\alph{equation}}
Eqs.(13) are easily solved to get: 
\begin{eqnarray}
\hat{B}_{+}(\omega)-\hat{B}_{-}^{+}(\omega) & = & \frac{(\omega^{2}+i\omega
\eta -\Omega^{2})\hat{A}_{0}(\omega)+i\omega\eta e^{\frac{i\omega a}{c}}
\hat{A}_{0}^{+}(-\omega)}{(\omega ^{2}+i\omega\eta -\Omega^{2}+i\omega\eta
e^{\frac{i\omega a}{c}})(\omega ^{2}+i\omega\eta -\Omega^{2}-i\omega\eta
e^{\frac{i\omega a}{c}} ) },\\
\hat{B}_{-}(\omega)-\hat{B}_{+}^{+}(\omega) & = & -\frac{i\omega\eta 
e^{\frac{i\omega a}{c}}
\hat{A}_{0}(\omega)+(\omega ^{2}+i\omega
\eta -\Omega ^{2})\hat{A}_{0}^{+}(-\omega)}{(\omega ^{2}+i\omega\eta -
\Omega ^{2}+i\omega\eta
e^{\frac{i\omega a}{c}})(\omega ^{2}+i\omega\eta -\Omega ^{2}-i\omega\eta
e^{\frac{i\omega a}{c} })}.
\end{eqnarray}
The roots of 
\def\theequation{\arabic{equation}}
\setcounter{equation}{14}
\begin{equation}     
\omega^{2}+i\omega \eta -\Omega^{2}+i\omega \eta e^{\frac{i\omega a}{c}}=0
\end{equation}
and 
\begin{equation}     
\omega^{2}+i\omega \eta -\Omega^{2}-i\omega \eta e^{\frac{i\omega a}{c}}=0
\end{equation}   
will determince the eigen decay rates and corresponding frequency shifts. 
These roots can not have positive imaginary part, namely the poles of $\hat{B}
_{+}(\omega)-\hat{B}_{-}^{+}(\omega)$ will not be localized in the upper
half $\omega$-plane, since two necessary conditions can be deduced for eq.(15)
to have root of positive imaginary part:
$$\eta >2\Omega \hspace{1cm} {\rm and} \hspace{1cm} \frac{\eta a}{c}>2\pi$$
and both of these conditions are untenable. Similarly eq.(16) also can not have
root of positive imaginary part. These results mean the basic physics 
laws will not be violated as in the case of monolayer$^{[11]}$.

We have derived four physical roots of eqs.(15) and (16) as:
\begin{eqnarray}
\omega_{1}=\Omega_{1}-i\Gamma_{1} & , & \hspace{0.5cm}
\omega_{3}=-\Omega_{1}-i\Gamma_{1}\nonumber\\
\omega_{2}=\Omega_{2}-i\Gamma_{2} & , & \hspace{0.5cm}
\omega_{4}=-\Omega_{2}-i\Gamma_{2}
\end{eqnarray}
in which 
\begin{eqnarray}
\Omega _{1}\cong \Omega (1-\frac{\eta^{2}}{2\Omega^{2}}+\frac{\eta a}{2c}) & , 
& \hspace{1.0cm}\Gamma_{1}\cong \eta,\nonumber\\
\Omega_{2}\cong \Omega (1-\frac{\eta a}{2c})\hspace{1.3cm} & , & \hspace{1.0cm}
\Gamma_{2}\cong \frac{1}{4}\eta\frac{\Omega ^{2}a^{2}}{c^{2}}.
\end{eqnarray}

In the following we will omit the terms proportional to $\hat{a}_{q}(0)$, 
$\hat{a}_{q}^{+}(0)$ as did in Ref[5], since here we just study the 
fluorescence of excitons.
Then the electric field is derived as follows$^{[11]}$
\setcounter{equation}{0}
\def\theequation{19\alph{equation}}
\begin{eqnarray}  
\hat{E} (z,t) & = & \frac{1}{2\pi}\int ^{\infty +i\epsilon}_{-\infty +i\epsilon}
 {\rm d}\omega \hat{E}(z,\omega) e^{-i\omega t}, \\
\hat{E}(z,\omega) & = & i\sum_{q}\sqrt{\frac{2\pi |q| c\hbar}{V}}
[\hat{a}_{q}(\omega)-\hat{a}_{-q}^{+}(\omega)]e^{iqz},  \\
\hat{a}_{q}(\omega)-\hat{a}_{-q}^{+}(\omega) & = & \frac{2\omega\sqrt{N}}{\Omega 
(\omega^{2}-q^{2}c^{2})}G(q)\sum _{k} {\Large O} (k-q) \nonumber \\
 & & \times [\omega (\hat{B}_{k}(\omega)-\hat{B}_{-k}^{+}(\omega))
 -i(\hat{B}_{k}(0)-\hat{B}_{-k}^{+}(0))]. 
\end{eqnarray}

The summation of $q$ in eq.(19b) can be tranformed to integration and carried 
out by contour integration. In the positive $z$ region outside the crystal 
slab, we get for the double lattice-layer case 
\def\theequation{\arabic{equation}}
\setcounter{equation}{19}
\begin{eqnarray}   
\hat{E}(z,\omega) &=& i\frac{2\pi \Omega d}{ac\sqrt{A}} [ \cos\frac
{\omega a}{2c} \frac{(\omega +\Omega)(\hat{B}_{+}(0)+\hat{B}_{-}(0))+
(\omega -\Omega)(\hat{B}^{+}_{+}(0)+\hat{B}^{+}_{-}(0))}
{\omega^{2}+i\omega \eta-\Omega^{2}+i\omega\eta 
e^{\frac{i\omega a}{c}}}\nonumber\\  & & +\sin\frac{\omega a}{2c}
\frac{(\omega +\Omega)(\hat{B}_{+}(0)-\hat{B}_{-}
(0))-(\omega -\Omega)
(\hat{B}^{+}_{+}(0)-\hat{B}^{+}_{-}(0))}{\omega^{2}+i\omega\eta -\Omega^{2}
-i\omega \eta e^{\frac{i\omega a}{c}}}]e^{i\frac{\omega}{c}z} 
\end{eqnarray}
where $A$ is the area of each layer, it is also the cross area of the 
normalization volume for the photon as mentioned above.

The electric field $\hat{E}(z,t)$ in this region is calculated by eq.(19a), 
with the results
\begin{mathletters}
\begin{eqnarray} 
\hat{E}(z,t)=0 & , & \hspace{0.5cm}for \hspace{0.5cm}z-ct>0 \\
\hat{E}(z,t)=\hat{\varepsilon}(z,t)+h.c.& , & \hspace{0.5cm}for \hspace{0.5cm}z-ct<0 
\end{eqnarray}
\end{mathletters}
where
\begin{eqnarray} 
\hat{\varepsilon}(z,t) & = & \frac{f}{c}\sqrt{\frac{\pi\hbar\Omega a}{8A}}
(1+\frac{\Omega}{\Omega_{1}}-i\frac{\Gamma_{1}}{\Omega_{1}})\cos\frac{
(\Omega_{1}-i\Gamma_{1})a}{2c}\nonumber\\
 & & [(\hat{B}_{+}(0)+\hat{B}_{-}(0))+\frac{\Omega_{1}-\Omega -i\Gamma_{1}}
 {\Omega_{1}+\Omega -i\Gamma_{1}}(\hat{B}_{+}^{+}(0)+\hat{B}_{-}^{+}(0))]
 e^{-i\Omega_{1}(t-\frac{z}{c})-\Gamma_{1}(t-\frac{z}{c})}\nonumber\\
 & + & \frac{f}{c}\sqrt{\frac{\pi\hbar\Omega a}{8A}}
(1+\frac{\Omega}{\Omega_{2}}-i\frac{\Gamma_{2}}{\Omega_{2}})\sin\frac{
(\Omega_{2}-i\Gamma_{2})a}{2c}\nonumber\\
 & & [(\hat{B}_{+}(0)-\hat{B}_{-}(0))-\frac{\Omega_{2}-\Omega -i\Gamma_{2}}
 {\Omega_{2}+\Omega -i\Gamma_{2}}(\hat{B}_{+}^{+}(0)-\hat{B}_{-}^{+}(0))]
 e^{-i\Omega_{2}(t-\frac{z}{c})-\Gamma_{2}(t-\frac{z}{c})}.
\end{eqnarray}

We note that this solution is free from Marckoffian approximation and also free
from antirotating wave interaction.

The electric field in the $z<0$ region can be derived similarly, with the 
resultant waves propagating in backward $z$ direction as expected.

There are two eigen decay rates appeared in the $\hat{E}(z,t)$: $\Gamma_{1}$ 
and $\Gamma_{2}$. The  corresponding eigen modes are linear combination of the
two modes of $m=\frac{1}{2}$ and $m=-\frac{1}{2}$.
As can be seen from eq.(20), these two eigen modes, which will be called as
superradiant mode and subradiant mode , correspond to the operators
$\frac{1}{\sqrt{2}}(\hat{B}_{+}(0)+\hat{B}_{-}(0))$ and
$\frac{1}{\sqrt{2}}(\hat{B}_{+}(0)-\hat{B}_{-}(0))$ respectively. Hence,
They correspond to modes of $k=0$ and $k=1$  with the operators
\setcounter{equation}{0}
\def\theequation{23\alph{equation}}
\begin{equation}  
\hat{B}_{0}(t)=\frac{1}{2}\sum _{l=\pm \frac{1}{2},k=\pm \frac{\pi}{2a}}
e^{ikal}\hat{B}_{k}(t)=\frac{1}{\sqrt{2}}[\hat{B}_{+}(t)+\hat{B}_{-}(t)],
\end{equation}
\begin{equation}
\hat{B}_{1}(t)=\frac{1}{2}\sum _{l=\pm \frac{1}{2},k=\pm \frac{\pi}{2a}}
e^{i(k-\frac{\pi}{a})la}\hat{B}_{k}(t)=\frac{1}{\sqrt{2}}[\hat{B}_{+}
(t)-\hat{B}_{-}(t)].
\end{equation}
Evidently, the dipoles of the two layers have the 
same phase for the former and have opposite phase for the latter. We note 
that the decay rate $\Gamma_{2}$ of subradiant mode here is still as large 
as $\frac{3\pi}{4}$ times the decay rate of a single atom(molecular), 
because the atoms in  each layer are still cooperated.
The decay rate of $k=0$ mode is twice of that of monolayer, which
is just the character of superfluorescence. As can be seen from eq.(22)
that even for the superradiant mode in which the emission is totally 
collective, the emitted light still may have different statistics and coherent
properties  according to the initial exciton state (also see the discussion 
in Ref[11]). For example, the coherent  part of 
the electric field $<\hat{E}(z,t)>$ will be nonzero  when the initial state 
of the exciton is a coherent state. But when the initial density matrix of the 
exciton is diagonal in Fock representation (including number state, chaotic
state), the coherent part of  $<\hat{E}(z,t)>$ will be zero.

Up to the first order of $\frac{\Omega a}{c}$ and $\frac{\eta}{\Omega}$, the 
$\hat{E}(z,t)$ is expressed by the superradiant mode operator 
$\hat{B}_{0}(0)$ and  the subradiant mode operator $\hat{B}_{1}(0)$ as follows:
\def\theequation{\arabic{equation}}
\setcounter{equation}{23}
\begin{eqnarray}
\hat{E}(z,t) & = & \sqrt{\frac{2\pi \eta \hbar \Omega}{cA}}
 [\hat{B}_{0}(0)-\frac{i\eta}{2\Omega}\hat{B}^{+}_{0}(0)]
 e^{-i\Omega _{1}(t-\frac{z}{c})-\Gamma_ {1}(t-\frac{z}{c})} \nonumber \\
 & & +\sqrt{\frac{2\pi \eta \hbar \Omega}{cA}}
(\frac{\Omega a}{2c}) \hat{B}_{1}(0)
e^{-i\Omega_{2}(t-\frac{z}{c})-\Gamma_{2}(t-\frac{z}{c})}+h.c.
\end{eqnarray}
for $z>0$ and $t-\frac{z}{c}>0$. Similar results for $z<0$, $t+\frac{z}{c}>0$.
$\hat{E}(z,t)$ is equal to zero if ($z>0$ , $t-\frac{z}{c}<0$) or
($z<0$, $t+\frac{z}{c}<0$)

Since we have seen irregular behavior in the usual intensity operator for the
solution of single layer case$^[11]$, here only the energy flux operator
 is given instead. The energy flux is usually defined by 
\begin{equation}
\hat{\bf S}(z,t)=\frac{c}{4\pi} :\hat{\bf E}(z,t)\times \hat{\bf B}(z,t):.
\end{equation}

It is readily to show that $\hat{\bf S}$ is always 
directed outward from the crystal film. So we rewrite $\hat{\bf S}$ as
 $\vec{n} \hat{S} $ which $\vec{n}$ is unit vector  directing outer space from
 lattice-layers. Namely,  
 it is  in positive $z$ direction in the $z>0$ region and in negative $z$ 
 direction in the $z<0$ region as required.

 So we may obtain $\hat{S}(z,t)$ from eq.(24).   After neglecting oscillating 
terms and higher order terms of $\frac{\eta}{\Omega}$ and 
$\frac{\Omega a}{c}$ (only keep first order terms), we have:
\begin{mathletters}
\begin{eqnarray}
\hat{S}(z,t)&=&\frac{\eta \hbar \Omega}{A}[\hat{B}^{+}_{0}(0)\hat{B}_{0}(0)
+\frac{i\eta}{2\Omega}\hat{B}^{2}_{0}(0)-
\frac{i\eta}{2\Omega}\hat{B}^{+2}_{0}(0)]e^{2\eta(\frac{z}{c}-t)}\nonumber \\
&+&\frac{\eta^{\prime}\hbar \Omega}{A}\hat{B}^{+}_{1}(0)\hat{B}_{1}(0)
e^{2\eta^{\prime}(\frac{z}{c}-t)} \nonumber \\
&+& \frac{\sqrt{\eta\eta^{\prime}}\hbar \Omega}{A}
[\hat{B}^{+}_{0}(0)\hat{B}_{1}(0)+\hat{B}^{+}_{1}(0)\hat{B}_{0}(0)\nonumber\\
&+&\frac{i\eta}{2\Omega}(\hat{B}^{+}_{0}(0)\hat{B}_{1}(0)+
\hat{B}_{0}(0)\hat{B}_{1}(0)  \nonumber \\
&-&\hat{B}^{+}_{1}(0)\hat{B}_{0}(0)-\hat{B}^{+}_{1}(0)\hat{B}^{+}_{0}(0))]
e^{(\eta+\eta^{\prime})(\frac{z}{c}-t)},
\end{eqnarray}
with
\begin{equation}
\eta^{\prime}=\eta\frac{\Omega^{2}a^{2}}{4c^{2}}.
\end{equation}
\end{mathletters}

We see from eq.(26) that the energy flux decays in three different rate.
The first term which is contributed by  the exciton of the short lifetime 
palys a important part  at the begining time. The second term  contributed by 
the  exciton of the long lifetime becomes dominat at late time. 
The  third term will exhibit itself in the intermediate stage.

\section{The case of triple lattice-layers}

The cases of odd $N$ and even $N$ have a qualitative difference in 
 the $m$-value series $-\frac{1}{2}(N-1),\cdots \frac{1}{2}(N-1)$ for eq.(4).
 In the former case, $m$ contains $zero$, while in the latter, not. 
 $N=3$ is the simplest case  of odd $N$, apart from the trival case 
 $N=1$, which has no nondiagonal terms $F_{mm'}$ (here and in the 
 following we use $F_{mm'}$ to denote $F_{kk'}$ according to the relation
 $k=\frac{2\pi m}{Na}$).Thus we  will study it as an example. 
 For $N=3$
 \begin{equation}   
 {\Large O} (k-q)=\frac{1}{3}[2\cos (k-q)a+1]
\end{equation}
leading to the matrix $F$ (with elements $F_{mm'}$,$\hspace{0.5cm} 
m, m'=1, 0, -1)$  as 
\begin{equation}  
F(\omega )=-\frac{iaf^{2}\Omega}{12\omega c}
\left (\begin{array}{ccc}
-x^{2}-2x+3 & -x^{2}+x     & 2x^{2}-2x\\
-x^{2}+x    & 2x^{2}+4x+3  & -x^{2}+x \\
2x^{2}-2x   & -x^{2}+x     & -x^{2}-2x+3
\end{array} \right ) \equiv -\frac{\eta \Omega}{6\omega} \eta D(\omega ),
\end{equation}
where $x=e^{\frac{i\omega a}{c}}\equiv e^{i\delta}$. To the second order of 
$\delta$, 
\begin{equation}  
D(\omega )=
\left (\begin{array}{ccc}
4\delta +3i\delta ^{2} & \delta +\frac{2}{3}i\delta ^{2} & 
-2\delta -3i\delta ^{2}\\
\delta +\frac{3}{2}i\delta ^{2} & 9i-8\delta -6i\delta ^{2} & 
\delta +\frac{3}{2}i\delta ^{2}\\
-2\delta -3i\delta ^{2} & \delta +\frac{3}{2}i\delta ^{2} & 
4\delta +3i\delta ^{2}
\end{array} \right ).
\end{equation}
We see that the nondiagonal elements are of the same order of $F_{11}$ and 
$F_{-1,-1}$, so that they can not be neglected in the equations for 
$\hat{B}_{1}-\hat{B}_{-1}^{+}$ and $\hat{B}_{-1}-\hat{B}_{1}^{+}$ 
($\hat{B}_{m}$ also means $\hat{B}_{k}$ for $k=\frac{2\pi m}{N a}$).

The couple equations now take the form
\begin{mathletters}
\begin{eqnarray}
&&(\omega ^{2}-\Omega ^{2})[\hat{B}_{m}(\omega )-\hat{B}_{-m}^{+}(\omega )]+
\frac{1}{3}\eta \omega \sum _{m'}D_{mm'}(\omega )[\hat{B}_{m'}(\omega )-\hat{B}
_{-m'}^{+}(\omega )] \nonumber \\
&&=i[(\omega +\Omega)\hat{B}_{m}(0)-(\omega -\Omega )\hat{B}_{-m}^{+}(0)]
+\frac{i}{3}\eta \sum _{m'}D_{mm'}(\omega )
[\hat{B}_{m'}(0)-\hat{B}_{-m'}^{+}(0)] 
\end{eqnarray}
with 
\begin{equation}
m, m^{\prime}=-1, 0, 1. 
\end{equation}
\end{mathletters}
In eqs.(30) the terms proportional to 
$\hat{a}_{q}(0)$ and $\hat{a}_{-q}^{+}(0)$ are neglected.

The eigen decay rates and corresponding frequency shifts are determined by the 
roots of the following equation
\begin{eqnarray}
&&(\omega ^{2}-\Omega ^{2})^{3}+\frac{1}{3}\eta \omega(B+2A)
(\omega ^{2}-\Omega ^{2})^{2}+\frac{1}{9}\eta ^{2}\omega ^{2}
(2AB+A^{2}-E^{2}-2C^{2})(\omega ^{2}-\Omega ^{2}) \nonumber \\
&&+\frac{1}{27}\eta^{3}\omega ^{3}[(A^{2}-E^{2})B+2C^{2}(E-A)]=0,
\end{eqnarray}
where $A,B,C$ and $E$ are matrix elements of $D(\omega )$, defined as follows:
\begin{equation}
D(\omega )=
\left (\begin{array}{ccc}
A(\omega) & C(\omega) & E(\omega)\\
C(\omega) & B(\omega) & C(\omega)\\
E(\omega) & C(\omega) & A(\omega)
\end{array} \right ).
\end{equation}
We get the six roots of eq.(31) as follows:
\begin{eqnarray}
\omega_{1} & = & \Omega_{1}-i\Gamma_{1} , \hspace{1.0cm}\omega_{1}'
=-\Omega_{1}-i\Gamma_{1},\nonumber\\
\omega_{0} & = & \Omega -i\Gamma , \hspace{1.3cm}
\omega_{0}'=-\Omega_{0}-i\Gamma_{0},\\
\omega_{-1} & = & \Omega_{-1}-i\Gamma _{-1} , 
\hspace{0.4cm}\omega_{-1}'=-\Omega _{-1}-i\Gamma_{-1}.\nonumber
\end{eqnarray}
with 
\setcounter{equation}{0}
\def\theequation{34\alph{equation}}
\begin{eqnarray} 
\Omega_{1}\cong\Omega(1-\frac{\eta a}{3c}) \hspace{1.5cm}& , & \hspace{0.5cm}
\Gamma_{1}=\frac{1}{27}\eta\frac{\Omega^{2}a^{2}}{c^{2}},\\
\Omega_{0}\cong\Omega (1-\frac{9\eta_{2}}{8\Omega^{2}}+
\frac{4\eta a}{3c}) &,& \hspace{0.5cm}\Gamma_{0}=\frac{3}{2}\eta,\\
\Omega_{-1}=(1-\frac{\eta a}{c}) \hspace{1.7cm}&,& \hspace{0.3cm}
\Gamma_{-1}=\eta\frac{\Omega^{2}a^{2}}{c^{2}}.
\end{eqnarray}
All the roots are in the lower half plan of complex $\omega$ as 
they should be. 

The direct way to solve for $\hat{B}_{m}(\omega)-\hat{B}_{-m}^{+}(\omega)$ from
eq.(30) is to diagonize the matrix $D(\omega)$ defined by eq.(28). Up to second 
order of $\delta$, we get the transformation matrix $T(\omega)$ which satisfies
\def\theequation{\arabic{equation}}
\setcounter{equation}{0}
\def\theequation{35\alph{equation}}
\begin{equation}  
T(\omega)D(\omega)\widetilde{T}(\omega)=
\left (\begin{array}{ccc}
D_{1}(\omega) & & \\
 & D_{0}(\omega) & \\
 & & D_{-1}(\omega)
 \end{array}\right )
\end{equation}
as 
\begin{equation} 
T(\omega)=\left (\begin{array}{ccc}
\frac{M}{\sqrt{2}} & \frac{-\sqrt{2} CM}{B-A-E} & \frac{M}{\sqrt{2}}\\
\frac{CM}{B-A-E} & M & \frac{CM}{B-A-E}\\
\frac{1}{\sqrt{2}} & 0 & -\frac{1}{\sqrt{2}}
\end{array} \right ), 
\end{equation}
where 
\begin{equation}
M=\frac{1}{\sqrt{1+\frac{2C^{2}}{(B-A-E)^{2}}}}. 
\end{equation}
The result for $\hat{B}_{m}(\omega)-\hat{B}_{-m}^{+}(\omega)$ is expressed then
by 
\def\theequation{\arabic{equation}}
\setcounter{equation}{0}
\def\theequation{36\alph{equation}}
$$
\hspace*{-11.5cm}\hat{B}_{m}(\omega)-\hat{B}_{-m}^{+}(\omega)=
$$
\begin{equation}
\sum _{m'}T_{m',m}\frac{6i}{\omega 
^{2}-\Omega ^{2}+\frac{1}{3}\eta\omega D_{m'}}[(\omega +\Omega +\frac{1}{3}
\eta D_{m'})\hat{\beta}_{m'}^{(1)}(\omega,0)-(\omega -\Omega +\frac{1}{3}\eta
D_{m'})\hat{\beta}_{m'}^{(2)}(\omega,0) 
\end{equation}
in which 
\begin{equation} 
\hat{\beta}_{m}^{(1)}(\omega,0)=\frac{1}{6}\sum _{m'}
T_{mm'}(\omega)\hat{B}_{m'}(0), \hspace{0.5cm}
\hat{\beta}_{m}^{(2)}(\omega,0)=\frac{1}{6}\sum _{m'}
T_{mm'}(\omega)\hat{B}_{-m'}^{+}(0) 
\end{equation}

Substituting eqs.(36) into eqs.(19) and carrying out the integrations, it 
finally yields
\def\theequation{\arabic{equation}}
\setcounter{equation}{0}
\def\theequation{37\alph{equation}}
$$\hat{a}_{q}(\omega)-\hat{a}_{-q}^{+}(\omega)=\frac{4\omega}
{\omega ^{2}-q^{2}c^{2}}G(q)\sum _{m,m'}
[2\cos (\frac{2\pi m}{3}-qa)+1]\times$$
\begin{equation} 
\frac{iT_{m'm}}{\omega ^{2}-\Omega ^{2}+\frac{1}{3}\eta \omega D_{m'}}
[(\Omega +\omega)\hat{\beta}^{(1)}_{m'}(\omega,0)-(\Omega-\omega)
\hat{\beta}_{m'}^{(2)}(\omega,0)], 
\end{equation}
$$\hat{E}(z,\omega)=\frac{f}{c}\sqrt{\frac{6\pi \hbar\Omega}{A}}\sum_{m,m'}
[2\cos (\frac{2\pi m}{3}-\frac{\omega a}{c})+1] \times$$
\begin{equation}
\frac{iT_{m'm}}{\omega ^{2}-\Omega ^{2}+\frac{1}{3}\eta \omega D_{m'}}
[(\Omega +\omega)\hat{\beta}^{(1)}_{m'}(\omega,0)-(\Omega-\omega)
\hat{\beta}_{m'}^{(2)}(\omega,0)]   
\end{equation}
and 
\begin{equation}
\hat{E}(z,t)=\sqrt{\frac{3\pi \eta\hbar\Omega}{cA}}{\Large \theta}
 (t-\frac{z}{c})
\sum _{m}\hat{\alpha}_{m}e^{-i\Omega_{m}(t-\frac{z}{c})-\Gamma_{m}
(t-\frac{z}{c})}+h.c.,
\end{equation}
for $z>0$, $t-\frac{z}{c}>0$, where
\begin{eqnarray} 
\hat{\alpha}_{m} & = & \frac{1}{\Omega_{m}}\sum _{m'}[2\cos (\frac{2\pi m'}{3}-
\frac{\omega _{m}a}{c})+1]T_{mm'}(\omega_{m})\nonumber\\
 & & [(\Omega + \omega_{m})\hat{\beta}^{(1)}_{m}(\omega _{m},0)- 
(\Omega - \omega_{m})\hat{\beta}^{(2)}_{m}(\omega _{m},0)].
\end{eqnarray}
 To the leading term, $\hat{\alpha}_{m}$ are given by 
\def\theequation{\arabic{equation}}
\setcounter{equation}{37}
\begin{eqnarray}
\hat{\alpha}_{1} & \cong & i\frac{\Omega a}{9c}[\hat{B}_{1}(0)+
\hat{B}_{-1}(0)],\nonumber \\
\hat{\alpha}_{0} & \cong & \hat{B}_{0}(0),\\
\hat{\alpha}_{-1} & \cong & \frac{\Omega a}{\sqrt{3}c}[\hat{B}_{1}(0)-
\hat{B}_{-1}(0)],\nonumber 
\end{eqnarray}
We see that $m=\pm 1$ modes are not of eigen decay rates. On the contrary,
 the eigen modes are nearly maximum mix of these two modes.
 
 In terms of the creation operator for an excitation in the {\it l}th layer$^{
 [3,11]}$, we have
 \begin{equation} 
\hat{B}_{k}(t)=\frac{1}{\sqrt{N}}\sum_{l=-\frac{1}{2}(N-1)}^{\frac{1}{2}(N-1)}
e^{-ikla}\hat{B}_{l}(t)
\end{equation}
Thus we have, denoting $\hat{B}_{k}$ by $\hat{B}_{m}$ as before, the three 
eigen modes. They are approximated as following 
\begin{mathletters}
\begin{equation}
\frac{1}{\sqrt{2}}[\hat{B}_{m=1}(0)+\hat{B}_{m=-1}(0)] =  \frac{1}{\sqrt{6}}
[-\hat{B}_{l=1}(0)+2\hat{B}_{l=0}(0)-\hat{B}_{l=-1}^{+}],
\end{equation}
\begin{equation}
\hat{B}_{m=0}  =  \frac{1}{\sqrt{3}}[\hat{B}_{l=1}(0)+\hat{B}_{l=0}(0)+
\hat{B}_{l=-1}(0)],
\end{equation}
\begin{equation}
\frac{1}{\sqrt{2}}[\hat{B}_{m=1}(0)-\hat{B}_{m=-1}(0)] = \frac{i}{\sqrt{2}}
[\hat{B}_{l=1}(0)-\hat{B}_{l=-1}(0)],
\end{equation}
while 
\begin{equation}
\hat{B}_{m=\pm 1}(0)=\frac{1}{\sqrt{3}}[(-\frac{1}{2}\mp \frac{\sqrt{3}}{2}i)
\hat{B}_{l=1}(0)+\hat{B}_{l=0}(0)+(-\frac{1}{2}\pm \frac{\sqrt{3}}{2}i)
\hat{B}_{l=-1}(0)].
\end{equation}
\end{mathletters}

We see the superradiant mode$(m=0)$ has a decay rate $2\Gamma_{0}=3\eta$ 
which is triplet of that for monolayer,showing the emission is  totally
cooperative. However, as in the double layer case, the statistical properties
of the light of this mode still may have different varieties which depend
on the initial state of the excitons. For the $z>a$ region, we may also
 give the energy flux of the case of triple lattice-layers according to 
eq.(25) and eq.(37) as following:
\begin{equation}
\langle \hat{S}(z,t) \rangle=\langle \hat{S}_{1}(z,t)\rangle+
\langle \hat{S}_{2}(z,t) \rangle,
\end{equation}
where $\langle \hat{S}_{1}(z,t)\rangle$ is the main part, it is expressed by
\begin{eqnarray}
\langle \hat{S}_{1}(z,t)\rangle
&=&\frac{\hbar\Omega\eta}{6A}
\{9\langle \hat{B}^{+}_{0}(0)\hat{B}_{0}(0) \rangle e^{-3\eta(t-\frac{z}{c})}
 \nonumber \\
&+&\frac{2\Omega^{2}a^{2}}{9c^{2}}
\langle \hat{B}^{+}_{+}(0)\hat{B}_{+}(0) \rangle
e^{-\frac{8\eta^{\prime}}{27}(t-\frac{z}{c})}  \nonumber\\
&+&\frac{6\Omega^{2}a^{2}}{c^{2}} 
\langle (\hat{B}^{+}_{-}(0)\hat{B}_{-}(0) \rangle
e^{-8\eta^{\prime}(t-\frac{z}{c})} \}
\end{eqnarray}
with $\hat{B}_{\pm}(0)=\frac{1}{\sqrt{2}}(\hat{B}_{1}(0) \pm \hat{B}_{-1}(0) )$ 
and $\langle \hat{S}_{2}(z,t) \rangle$ may be approximated by
\begin{eqnarray}
\langle \hat{S}_{2}(z,t) \rangle &=& 
-i\frac{\hbar \Omega}{3\sqrt{3}A}\sqrt{\eta\eta^{\prime}}
[\langle \hat{B}^{+}_{+}(0) \hat{B}_{0}(0)-
\hat{B}^{+}_{0}(0) \hat{B}_{+}(0)\rangle + \nonumber \\
& & i3\sqrt{3}
\langle \hat{B}^{+}_{-}(0) \hat{B}_{0}(0)-
\hat{B}^{+}_{0}(0) \hat{B}_{-}(0)\rangle]e^{-\frac{3}{2}\eta(t-\frac{Z}{c})}.
\end{eqnarray}
It only appears when the initial exciton density matrix in Fock representation
has non-diagonal elements.

\section{Brief summary}
${\bf 1.}$  Knoester$^{[3]}$ claimed that $F_{kk'}(\omega)$ 
is diagonal to a good approximation. But we show explicitly that this is not 
generally true.

${\bf 2.}$  In low-density case, even the emission is superradiant
in nature, the light still may have different coherent statistical
properties, depending on the initial state of exciton.

\vspace*{10pt}
\centerline{\large \bf Acknowledgement}
This work is a part of the project 19774004 supported by National Science 
Foundation of China. It is also partially supported by the international
 program of the National Science Foundation of USA (INT-9974051).
\newpage  
\centerline{\large \bf References}
\begin{enumerate}
\item J.J.Hopfield, {\em Phys. Rev.} {\bf 112}, 1555(1958)  
\item V.M.Aranvovich,   {\em Sov. Phys. JETP }{\bf 10}, 
      307(1960)  
\item E.Hanamura, {\em Phys. Rev.} B {\bf 38}, 1228(1988) 
\item L.C.Andreani, F.tassone and F.Bassani, {\em solid state commun.} 
      {\bf 77}, 641(1991) 
\item J.Knoester,  {\em Phys. Rev. Lett.}{\bf 68}, 654(1992)  
\item D.S.Citrin,  {\em Phys. Rev.} B {\bf 47},  3832(1993)   
\item T.Tokihiro, Y.Manabe and E.Hanamura,  {\em Phys.Rev.} 
      B {\bf 47},  2019(1993)
\item G.Bj\"{o}rk,  S.Pau,  J.Jacobson and Y.Yamamoto,  {\em Phys. Rev.}
       B {\bf 50},  17336(1994)
\item G.Bj\"{o}rk,  S.Pau,  J.Jacobson, H.Cao and Y.Yamamoto,  
       {\em Phys. Rev.} B {\bf 52},  17310(1995)         
\item T.Tokihiro, Y.Manabe and E.Hanamura,  {\em Phys.Rev.} 
      B {\bf 51},  7655(1995)
\item Chang-qi Cao, Hui Cao and F. Haake,  {\em to be published}
\end{enumerate}
\end{document}